\documentclass{rnaastex}

\begin{document}

\title{On Distinguishing Interstellar Objects Like `Oumuamua From Products of Solar System Scattering}

%% Note that the corresponding author command and emails has to come
%% before everything else. Also place all the emails in the \email
%% command instead of using multiple \email calls.
\correspondingauthor{Jason T.\ Wright}
\email{astrowright@gmail.com}

\author[0000-0001-6160-5888]{Jason T.\ Wright}
\affil{Department of Astronomy \& Astrophysics / 525 Davey Laboratory /
The Pennsylvania State University /
University Park, PA, 16802, USA}
\affil{Center for Exoplanets and Habitable Worlds / 525 Davey Laboratory /
The Pennsylvania State University /
University Park, PA, 16802, USA}

%% Note that RNAAS manuscripts DO NOT have abstracts.
\keywords{celestial mechanics---Oort Cloud---Kuiper belt: general---minor planets, asteroids: individual (1I/2017 U1 (`Oumuamua))}

%% Start the main body of the article. If no sections in the 
%% research note leave the \section call blank to make the title.
\section{Could `Oumuamua Possibly Be a Solar System Object?} 
The recent discovery of the apparently interstellar\footnote{\url{https://www.minorplanetcenter.net/mpec/K17/K17UI1.html}} asteroid 1I/2017 U1 (`Oumuamua)\footnote{\url{https://www.minorplanetcenter.net/mpec/K17/K17V17.html}} has led to speculation about its origin. \citep{Schneider18} explored the possibility that it is a Solar System object scattered from a Solar System planet and concludes that none of the known planets could be the scatterer, nor could the hypothetical ``Planet Nine'' proposed by \citet{Batygin16}.  \citet{Schneider18} concludes that if `Oumuamua is a Solar System object, it must have been scattered by ``another, yet unknown planet.''

If this were true, then one might search for such a planet in the directions of the incoming trajectory of `Oumuamua and similar objects discovered in the future. However, the orbital energy of `Oumuamua is too large to be the result of a single scattering event from {\it any} hypothetical Solar System object. 

\section{An Upper Limit on the Distance of the Hypothetical Scattering Event}

\citet{Zwart17} give an empirical discussion from Monte Carlo simulations on why 'Oumuamua is unlikely to be the product of an encounter between objects bound to the Sun.

Here, I derive an extremely conservative upper limit on the total orbital energy of a scattered object in the Solar System by approximating the gravitational scattering event as a two-body elastic collision.

Let $v_i$ and $V_i$ be the initial, pre-interaction speeds of two objects of mass $m$ and $M$ respectively in the heliocentric frame, and that $M\gg m$. In such a two-body collision, the maximum final velocity of the lighter object occurs for a head-on (impact parameter $b = 0$) encounter, which  results in a final velocity for the scattered object of $v_f = v_i + 2V_i$ (in the opposite direction from its initial velocity).  For any larger impact parameter (which, in practice, has a lower limit of the radius of the scatterer), we have 
\begin{equation}
v_f < v_i + 2V_i
\end{equation}
By assumption, both objects are bound to the Sun before the interaction, and so must have 
\begin{eqnarray}
v_i < \sqrt{2}v_{\rm circ}\\
V_i < \sqrt{2}v_{\rm circ}
\end{eqnarray}
\noindent where 
\begin{equation}
v_{\rm circ} \equiv \sqrt{GM_\odot/r}
\end{equation}
\noindent is the circular orbital velocity about the Sun at the heliocentric distance of the scattering event, $r$. 

This yields $v_f < 3\sqrt{2} v_{\rm circ}$.  The total energy of the scattered object after the encounter is then 
\begin{equation}
E = \frac{1}{2}mv_f^2 - \frac{GM_\odot}{r} < 8mv_{\rm circ}^2
\end{equation}
\noindent or, expressed as the (now unbound) scattered object's velocity very far from the Sun, $v_\infty$, we have
\begin{equation}
v_\infty < 4 v_{\rm circ}
\end{equation}
\noindent or, expressed in terms of the heliocentric distance of the scattering event,
\begin{equation}
r < 16 G M_\odot / v_\infty^2 = 1.4\times 10^4 {\rm AU} \left(\frac{v_\infty}{\rm{km/s}}\right)^{-2}
\end{equation}
This analysis applies to bound scatterers of arbitrarily high escape velocity, but assumptions about the radius and density of the scatterer can provide an even tighter constraint, which is a subject for future refinement of this limit. 

\section{The Case of `Oumuamua and Future Interstellar Visitors}

For the case of `Oumuamua, \citet{FuenteMarcos18} give orbital elements consistent with those 
\citet{Gaidos17} and \citet{Mamajek17} quote from the JPL Small-Body Database Browser\footnote{which I confirmed are still valid on 27 November 2017 at \url{https://ssd.jpl.nasa.gov/sbdb.cgi?sstr=A\%2F2017\%20U1}}, which yield $v_\infty = 26.2 \pm 0.1$ km/s.  The limits described above show that any potential Solar System scatterer must have had orbital velocity higher than any bound object can have beyond 21 AU from the Sun.  Since our census of Solar System objects capable of scattering `Oumuamua is presumably complete within this limit, no  hypothetical Solar System object could have scattered a previously bound object onto the present trajectory of `Oumuamua.  Indeed, this limit makes clear that typical interstellar objects and objects scattered in from the outer Solar System are easily distinguished by their $v_\infty$ values, which differ by at least an order of magnitude.

This limit alone does not prove beyond all doubt that `Oumuamua is of interstellar origin: a more complex scattering history is not excluded by this analysis, nor is a scattering event off of an unbound object such as a passing brown dwarf. One might also consider whether fast-moving fragments can be generated from collisions in the outer Solar System.

However, \citet{Mamajek17} demonstrates that the kinematics of `Oumuamua are exactly what ``might be expected of interstellar field objects,'' and indeed its $v_\infty$ is consistent with predictions by \citet{Cook16}, so its origins in interstellar space seem all but certain (although \citet{Schneider18}  cleverly points out the (extraordinarily unlikely) possibility that it could be a {\it former} Solar System object returning from an interstellar voyage.) 

Nonetheless, the limit described above---or, even better, tighter refinements of this limit in terms of the scatterer's escape velocity---can be used to reject the possibility that objects detected in the future on hyperbolic orbits might be the result of a scattering from an as-yet-undetected outer Solar System planet; or to verify that searches for an such a planet might be profitable.

\acknowledgments

I thank Eric Ford, Steinn Sigurdsson, James Davenport, and Andrew Shannon for helpful conversations.

\bibliography{Oumuamua}

\begin{thebibliography}{}
\expandafter\ifx\csname natexlab\endcsname\relax\def\natexlab#1{#1}\fi
\providecommand{\url}[1]{\href{#1}{#1}}

\bibitem[{Batygin \& Brown(2016)}]{Batygin16}
Batygin, K., \& Brown, M.~E. 2016, The Astrophysical Journal Letters, 833, L3.
\newblock \url{http://stacks.iop.org/2041-8205/833/i=1/a=L3}

\bibitem[{Cook {et~al.}(2016)Cook, Ragozzine, Granvik, \& Stephens}]{Cook16}
Cook, N.~V., Ragozzine, D., Granvik, M., \& Stephens, D.~C. 2016, The
  Astrophysical Journal, 825, 51.
\newblock \url{http://stacks.iop.org/0004-637X/825/i=1/a=51}

\bibitem[{de~la Fuente~Marcos \& de~la Fuente~Marcos(2017)}]{FuenteMarcos18}
de~la Fuente~Marcos, C., \& de~la Fuente~Marcos, R. 2017, Research Notes of the
  AAS, 1, 5.
\newblock \url{http://stacks.iop.org/2515-5172/1/i=1/a=5}

\bibitem[{Gaidos {et~al.}(2017)Gaidos, Williams, \& Kraus}]{Gaidos17}
Gaidos, E., Williams, J., \& Kraus, A. 2017, Research Notes of the AAS, 1, 13.
\newblock \url{http://stacks.iop.org/2515-5172/1/i=1/a=13}

\bibitem[{Mamajek(2017)}]{Mamajek17}
Mamajek, E. 2017, Research Notes of the AAS, 1, 21.
\newblock \url{http://stacks.iop.org/2515-5172/1/i=1/a=21}

\bibitem[{{Portegies Zwart} {et~al.}(2017){Portegies Zwart}, {Pelupessy},
  {Bedorf}, {Cai}, \& {Torres}}]{Zwart17}
{Portegies Zwart}, S., {Pelupessy}, I., {Bedorf}, J., {Cai}, M., \& {Torres},
  S. 2017, ArXiv e-prints, arXiv:1711.03558

\bibitem[{Schneider(2017)}]{Schneider18}
Schneider, J. 2017, Research Notes of the AAS, 1, 18.
\newblock \url{http://stacks.iop.org/2515-5172/1/i=1/a=18}

\end{thebibliography}

\end{document}